\documentclass[a4paper]{jpconf}
\usepackage{graphicx}
\begin{document}

\title{Mechanical losses in low loss materials studied by Cryogenic Resonant Acoustic spectroscopy of bulk materials (CRA spectroscopy)}

\author{Anja Zimmer$^1$, Ronny Nawrodt$^1$, Daniel Heinert$^1$, Christian Schwarz$^1$, Matthias Hudl$^1$, Torsten Koettig$^1$, Wolfgang Vodel$^1$, Andreas T\"unnermann$^2$ and Paul Seidel$^1$}

\address{$^1$ Institute of Solid State Physics, Friedrich-Schiller-University Jena, Helmholtzweg~5, D-07743~Jena, Germany}
\address{$^2$ Institute of Applied Physics, Friedrich-Schiller-University Jena, Max-Wien-Platz~1, D-07743~Jena, Germany}

\ead{anja.zimmer@uni-jena.de}

\begin{abstract}
Mechanical losses of crystalline silicon and calcium fluoride have been analyzed in the temperature range from 5 to 300~K by our novel mechanical spectroscopy method, cryogenic resonant acoustic spectroscopy of bulk materials (CRA spectrocopy). The focus lies on the interpretation of the measured data according to phonon-phonon interactions and defect induced losses in consideration of the excited mode shape.
\end{abstract}

\section{Introduction}
Low mechanical loss materials are demanded in instruments with extraordinary precision to decrease their thermal noise as mechanical losses and thermal noise are linked \cite{Cal1951}. For instance, low loss materials like crystalline silicon, sapphire, calcium fluoride and quartz are candidates for the optical components of future cryogenic interferometric gravitational wave detectors \cite{Row2005}. Due to the anisotropy of the crystals and the search for an optimal operating temperature of the detectors systematic measurements of the mechanical losses have to be performed. Therefore, our novel mechanical spectroscopy method, cryogenic resonant acoustic spectroscopy of bulk materials (CRA spectrocopy), was applied in the temperature range from 5 to 300~K. Instead of the small losses in the detection band, the reciprocals of the losses at the resonant frequencies, the mechanical quality factors Q, have been measured. Following the test of our method on the fairly well-known material crystalline quartz \cite{Zim2007}, results on silicon and calcium fluoride are presented. For details on the measuring set-up and method see references \cite{Naw2006} and \cite{Hei2007}.

\section{Results on crystalline silicon (100)}
\begin{figure}[htb]
a)~\includegraphics[width=.46\textwidth]{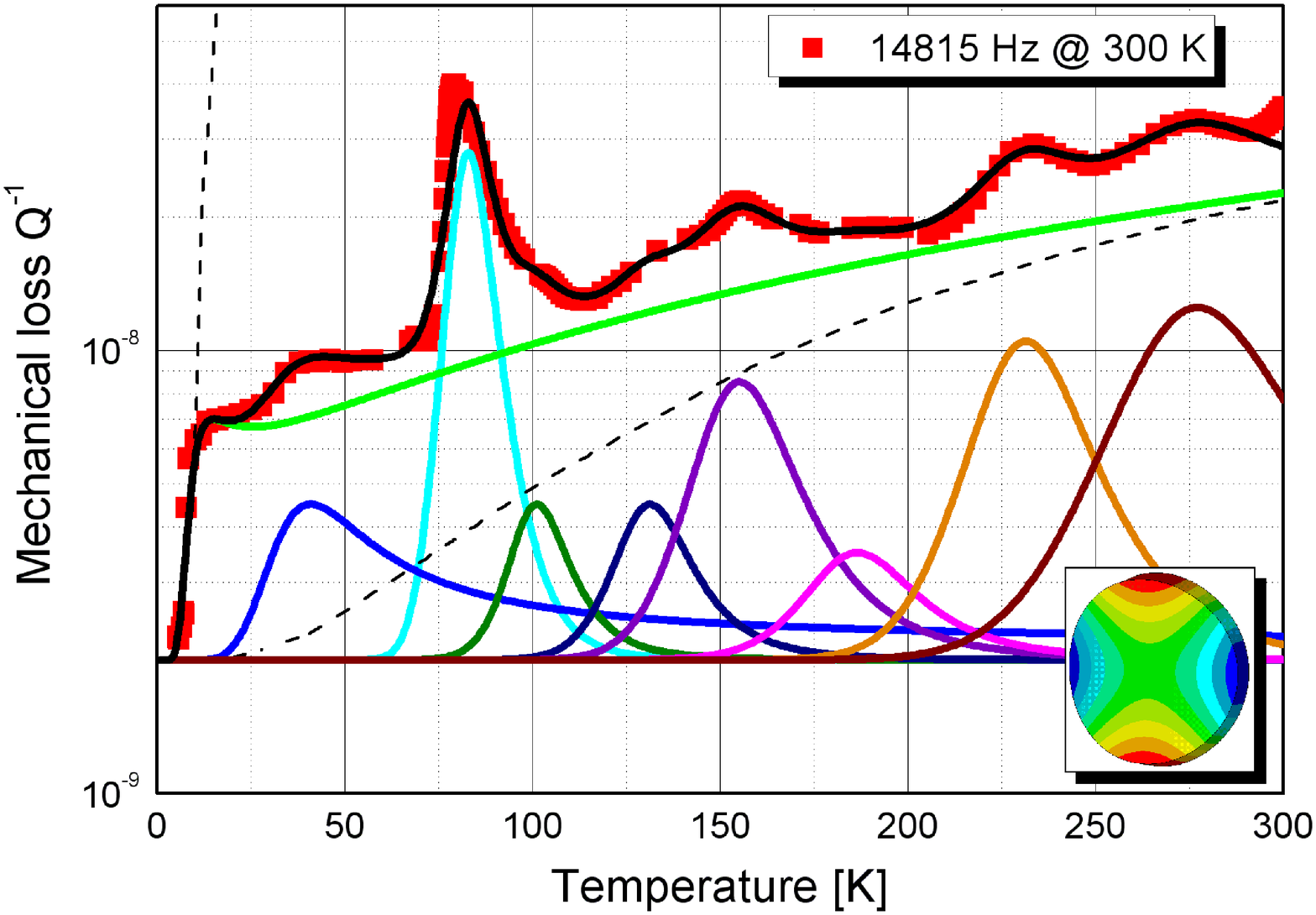}
b)~\includegraphics[width=.46\textwidth]{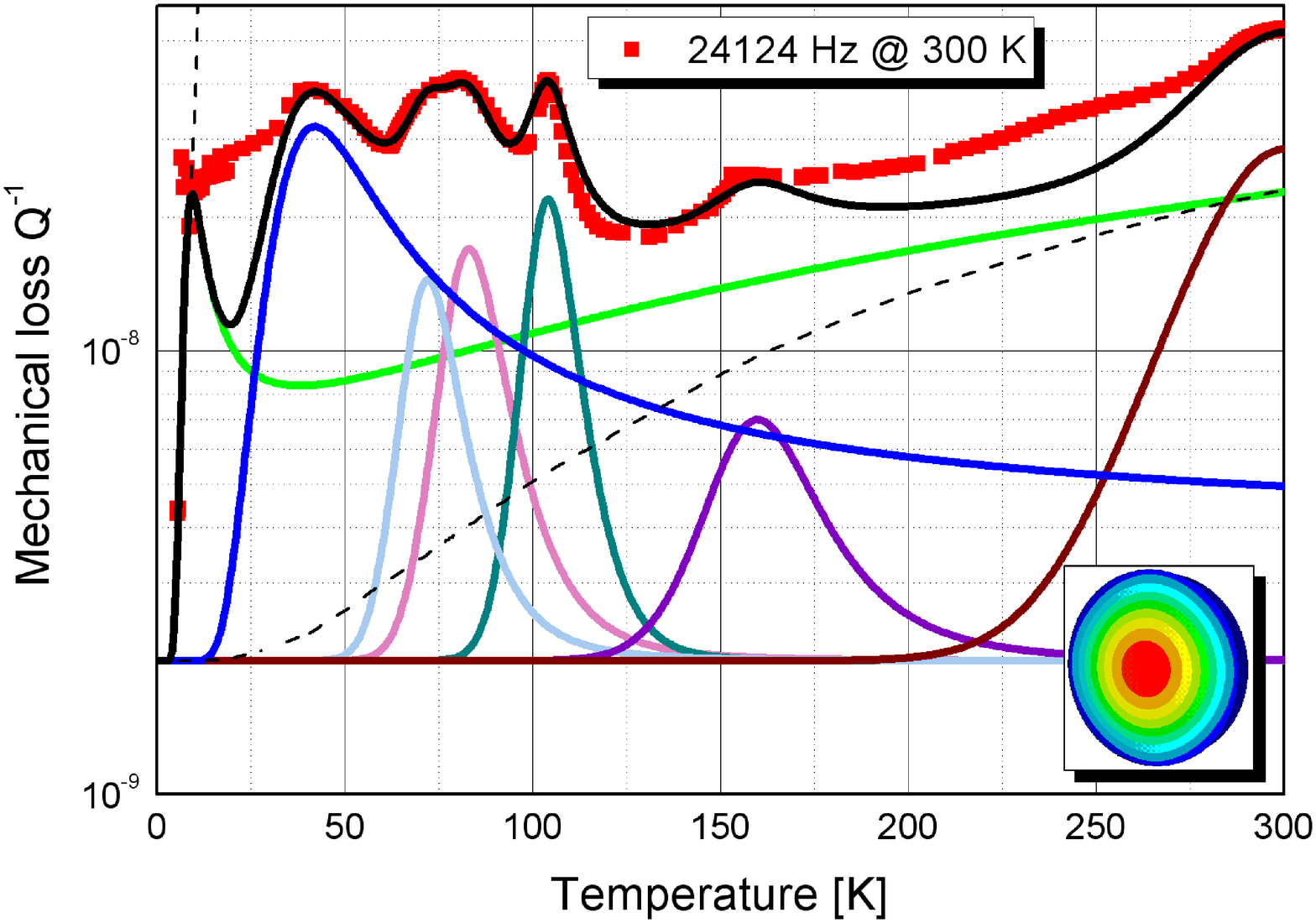}
\caption{Damping versus temperature gained from Q measurements on a silicon sample. Red squares: measured data. For fitting parameters see text. Insets show excited mode shapes. Colours indicate displacement in direction of the cylinder axis. Blue and red: maximum, green: minimum.}
\label{fig:si}
\end{figure}

The sample was of cylindrical shape, 76.2~mm in diameter and 12~mm thick. The reciprocals of the measured Q factors, corresponding to the mechanical loss at the particular resonant frequency and temperature, of two selected modes have been plotted in Fig.\,\ref{fig:si}\,a) and b).
\begin{table}
\centering
		\begin{tabular}{|l|c|r|r|r|r|c|}
		  \hline
			\textsc{relaxation} & \textsc{frequency} & \textsc{T$_{peak}$} & \textsc{$\Delta_{0}$} & \textsc{$\tau_{0}$} (s) & \textsc{$E_{a}$} & $d\left(\vartheta,f_{r}\right)$ \\
			\textsc{peak} & \textsc{@ 300 K} (Hz) & (K) & ($\times10^{-10}$) & ($\times10^{-15}$) & (meV) & \\
			\hline
  		1. (light green)	& 14815	 & 15 	& 7.8		&	92$\times10^{7}$ 	&	2.5 & $\vartheta$\\
  											& 24124	 & 9 	&	45		&	9$\times10^{7}$	&	3.3 & $\vartheta$\\
  		\hline
  		2. (blue) 	& 14815  & 40 	&	50 	&	3.8$\times10^{8}$ 				&	12 & 1\\
  		 						& 24124  & 41 	&	600 	&	2$\times10^{8}$ 				&	12.7 & 1\\
  		\hline
			3. (light cyan)	& 24124  & 72 	& 250 		&	9$\times10^{3}$ 				&	84 & 1\\		
			\hline
			4. (light magenta)	& 24124 	& 83 &	300 		&	5$\times10^{3}$ 				&	101 & 1\\		
			\hline
			5. (cyan)				& 14815 	& 83 &	520		&	35					&	140  & 1\\
			\hline
			6. (dark green)	& 14815 	& 100 	&	50 		&	12$\times10^{2}$ 			&	140 & 1\\
			\hline
			7. (dark cyan)	& 24124 	& 104 &	400 		&	1 				&	203 & 1\\		
			\hline
			8. (dark blue)	& 14815 	& 131 	& 50 		&	12$\times10^{2}$ 				&	185 & 1\\
			\hline
			9. (violet)				& 14815 	& 155 	& 130 		&	25$\times10^{2}$ 				&	205 & 1\\
												& 24124 	& 160 	& 100 		&	30$\times10^{2}$ 				&	200 & 1\\
			\hline
			10. (magenta)	& 14815 	& 185 	& 30 		&	2$\times10^{3}$					&	250 & 1\\
			\hline
			11. (orange)	& 14815 	& 230 	& 170 		&	20 										&	402 & 1\\
			\hline
			12. (brown)				& 14815 	& 278 	& 210 		&	14$\times10^{2}$ 				&	380 & 1\\
												& 24124 	& 		 	& 530 		&	28$\times10^{2}$ 				&	380 & 1\\
			\hline
		\end{tabular}
		\caption{Fit parameters of the relaxation peaks in Fig.\,\ref{fig:si}\,a) and b). The upper limit of the background damping is  $2\times10^{-9}$.}
		\label{tab:parsi}
\end{table}
According to the superposition principle of mechanical losses, the total damping curve has been decomposed into contributions (displayed by coloured lines) due to single anelastic processes of different origin. The loss $\phi$ caused by a single relaxation process with small relaxation strength $\Delta$ at frequency f has been modelled by \cite{deB1972}
\begin{equation}
        \phi  = \Delta \; 2\pi\,f\,\tau / \left( 1+ \left( 2\pi\,f\,\tau \right)^{2} \right), \  \Delta = \Delta_{0}\;d\left(\vartheta,f_{r}\right),
        \label{loss}
\end{equation}
where $\tau$ is the relaxation time, $\Delta_{0}$ is a dimensionless coupling strength and $d\left(\vartheta,f_{r}\right)$ is in general a dimensionless function of reduced temperature $\vartheta$ and frequency $f_{r}$. The maximum damping occurs when $2\pi\,f\,\tau=1$. In Q measurements f is the resonant frequency specified by the material, sample geometry and temperature T, whereas $\tau$ is characteristic for the loss process and depends on T. The relaxation strength may vary with f and T and furthermore depends on the excited mode shape. Therefore, one aim is to extrapolate the relaxation strength in the off-resonant region by performing systematic Q measurements.
The relaxation time was assumed to follow an Arrhenius-like law \cite{deB1972}
\begin{equation}
        \tau  =  \tau_{0}\,exp\left(E_{a} / k_{B}\,T\right),
\label{arrh}
\end{equation}
with the relaxation constant $\tau_{0}$, activation energy $E_{a}$, and Boltzmann constant $k_{B}$. The parameters of the fitted relaxation peaks are arranged in Tab.\,\ref{tab:parsi}. The mechanical losses are dominated over the whole temperature range by dissipation processes assigned to interactions of the acoustic waves with thermal phonons. Landau and Rumer \cite{Lan1937} developed a theoretical description for the lower temperature region of the peak ($2\pi\,f>1$) whereas Akhieser \cite{Akh1939} treated the higher temperature region ($2\pi\,f<1$). The results gained by applying their theory are indicated by dashed black lines in Fig.\,\ref{fig:si} a) and b). Nevertheless, the 'phonon-peak' is well approximated by assuming a relaxation strength proportional to T and relaxation times according to Eq.\,\ref{arrh}. For the mode in Fig.\,\ref{fig:si} a) the result is better than for that in Fig.\,\ref{fig:si}\,b). The quality of the description of the phonon-phonon interactions is of outstanding relevancy as the quality of the fit of the remaining relaxation peaks strongly depends on it. Several parameters of the remaining peaks have been related to formerly observed reorientation processes. 
Lam and Douglass \cite{Lam1981} reported an activation energy of $E_{a}$$=$0.14~eV and relaxation constants of $\tau_{0}$$=$ $7.2\times10^{-12}$s resp. $\tau_{0}$$=$$4.7\times10^{-12}$s in loss measurements which could be assigned to vibrations of Si-O-Si complexes by IR measurements on oxygen-doped silicon. The dark green peak in Fig.\,\ref{fig:si}\,a) shows similar parameters.
Lam and Douglass further related the parameters $E_{a}$$=$0.20~eV and $\tau_{0}$$=$$8.64\times10^{-13}$s to the oxygen-vacancy (O-V) complex. Watkins and Corbett \cite{Wat1961} had at that time already observed by ESR that different kinds of redistribution can take place in an O-V center. An electronic redistribution is related to the mentioned parameters whereas a defect reorientation is characterized by $E_{a}$$=$0.38~eV and a relaxation time of $\tau_{0}$$=$$2\times10^{-13}$s which had been also observed by Berry \cite{Ber1979}. In the damping curves of both modes the violet and brown peaks might be of that origin. Coutinho et al. \cite{Cou2003} investigated a hydrogen atom jumping between two Si dangling bonds in vacancy-oxygen-hydrogen complexes (VOH and VOH$_{2}$ defects) with $E_{a}$$=$0.18~eV. The related relaxation constant reported by Johannesen et al. \cite{Joh2000} is $\tau_{0}$$=$$1.1\times10^{-12}$s. The dark blue relaxation peak in Fig.\,\ref{fig:si}\,a) is probably caused by this defect. As the O-V as well as the VOH complex anneals out at 573~K this sample preparation might decrease the damping in the corresponding temperature regions. According to the attenuation measurements of Pomerantz \cite{Pom1970} the sample is presumably doped with phosphor (blue peak). Regarding the relaxation strengths the dependence on mode shape is especially visible in the low temperature region.
\begin{figure}[htb]
a)~\includegraphics[width=.46\textwidth]{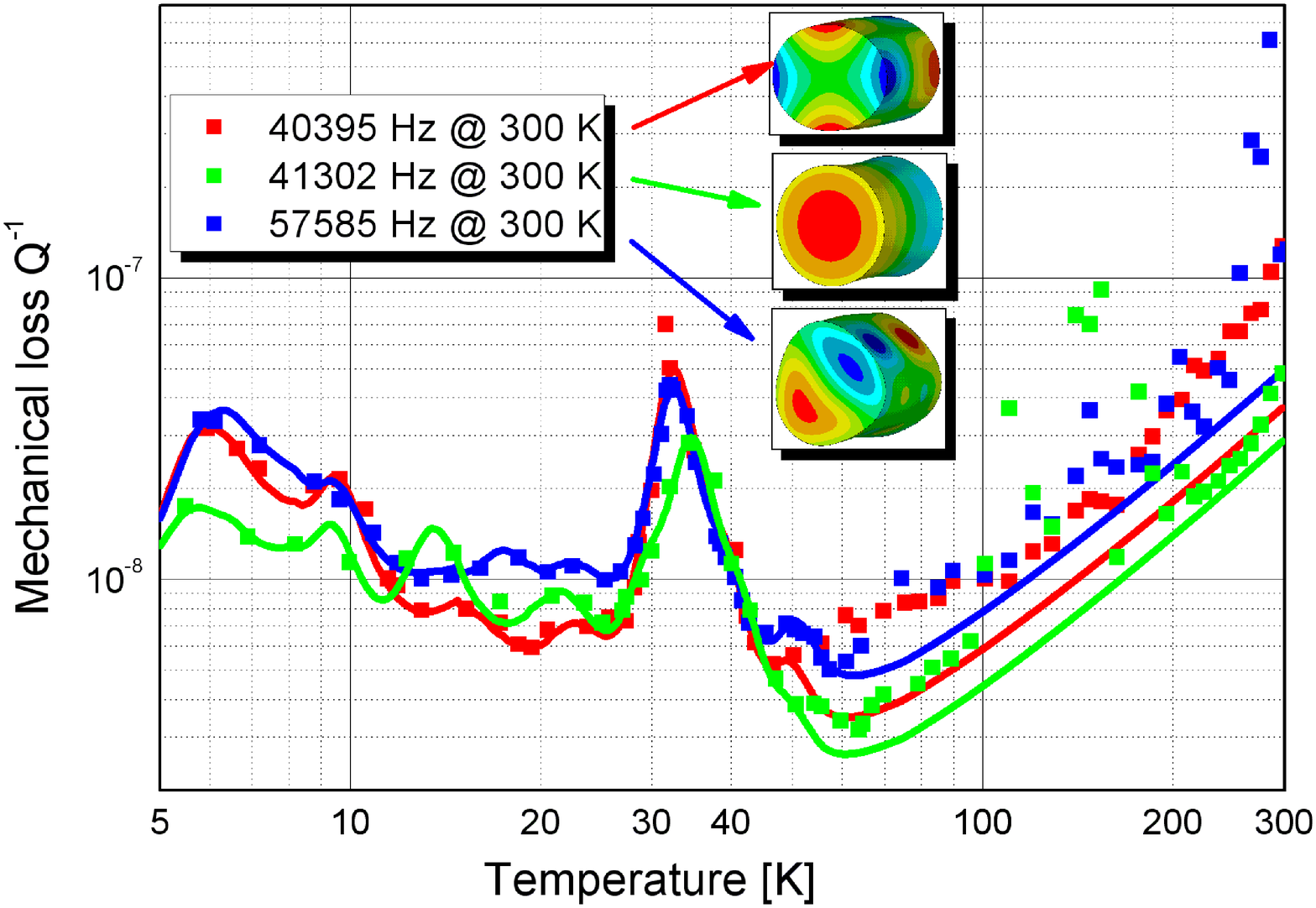}
b)~\includegraphics[width=.46\textwidth]{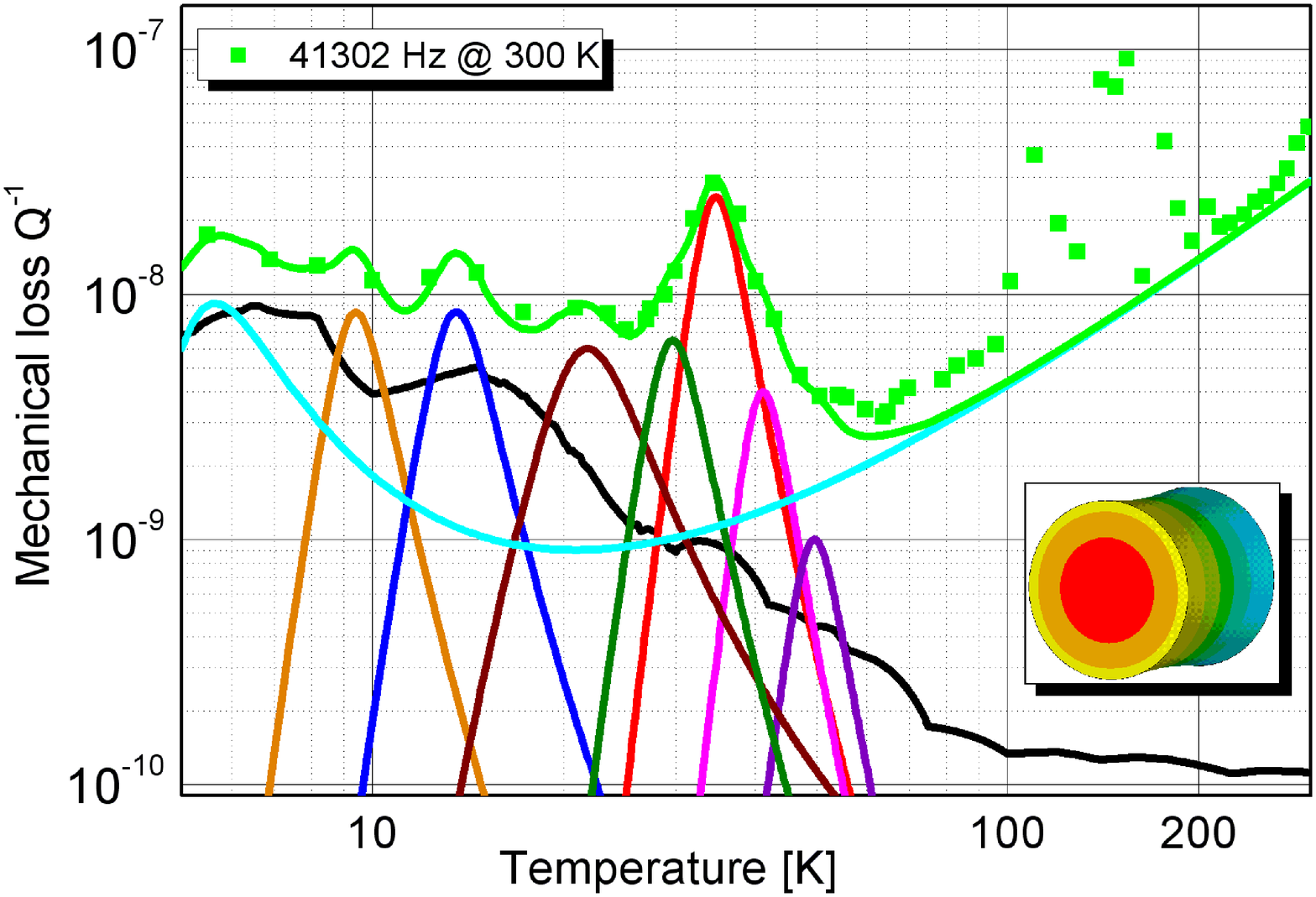}
\caption{Damping gained from Q measurements on calcium fluoride sample. For details see text.}
\label{fig:caf2}
\end{figure}

\section{Results on crystalline calcium fluoride (100)}
Examining the mechanical losses of the selected modes of a sample made of CaF$_{2}$ (cylindrical shape, 75~mm in diameter, 75~mm thick) in Fig.\,\ref{fig:caf2}\,a) the variation with mode shape is far less existent as for silicon. Peaks with the same characteristic parameters appear in all damping curves except very narrow loss peaks above 100~K due to coupling of the substrate motion to the suspension \cite{Naw2007}. Therefore, the results on one mode have been examplarily plotted in Fig.\,\ref{fig:caf2}\,b) with corresponding fitting parameters in Tab.\,\ref{tab:parcaf2}. 
\begin{table}
\centering
		\begin{tabular}{|l|r|r|r|r|c|}
		  \hline
			\textsc{relaxation} &  \textsc{T$_{peak}$} & \textsc{$\Delta_{0}$} & \textsc{$\tau_{0}$} (s) & \textsc{$E_{a}$} & $d\left(\vartheta,f_{r}\right)$\\
			\textsc{peak} &  (K) & ($\times10^{-10}$) & ($\times10^{-13}$) & (meV) & \\
			\hline
  		1. (cyan)	&  5 	& 6 		&	1.9$\times10^{4}$ 	&	3.6 & $\vartheta^{2}$ \\
  		\hline
  		2. (orange) 	  & 9 	&	170 	&	30 				&	11.4 & 1\\
  		\hline
			3. (blue)	  & 13 	& 170 		&	1$\times10^{2}$ 				&	15 & 1\\		
			\hline
			4. (brown)	 	& 21 &	120 		&	1$\times10^{4}$ 				&	15.5 & 1\\		
			\hline
			5. (dark green)	 	& 29 &	130		&	20					&	37 & 1 \\
			\hline
			6. (red)	 	& 34 	&	500 		&	3  			&	49 & 1\\
			\hline
			7. (magenta)	 	& 41 &	80 		&	1 				&	62 & 1\\		
			\hline
			8. (violet)	 	& 49 	& 20 		&	3 				&	70 & 1\\
			\hline
		\end{tabular}
		\caption{Fit parameters of the damping peaks shown in Fig.\,\ref{fig:caf2}\,a).}
		\label{tab:parcaf2}
\end{table}
In the low temperature range the mechanical losses are dominated by thermoelastic damping (black line in Fig.\,\ref{fig:caf2}\,b)) \cite{Brag1985}
\begin{equation}
\phi_{te} = \kappa\,T\,\alpha^2\,\rho\,2\pi\,f / \left(9\,C^2\right),
\label{tedamping}
\end{equation}
with the thermal conductivity $\kappa$, the volume thermal expansion coefficient $\alpha$, density $\rho$ and specific heat C. Hence, the 'phonon-peak' is less visible. The damping at higher temperatures offers an orientation for fitting. According to it, the relaxation strength has been modelled to be proportional to T$^{2}$. An investigation of the losses below 5~K where the thermoelastic damping decreases would be helpfull for a better determination of the shape of the maximum. A further interpretation of the damping peaks according to their characteristic parameters is in progress.

\ack We thank F. Bechstedt for the helpfull discussion as well as the German DFG for the support under contract SFB Transregio 7.

\end{document}